\documentclass[12pt,epsf]{article}
\usepackage{graphicx}
\usepackage{epsfig}
\begin{document}

\def\a{\alpha}
\def\b{\beta}
\def\c{\varepsilon}
\def\d{\delta}
\def\e{\epsilon}
\def\f{\phi}
\def\g{\gamma}
\def\h{\theta}
\def\k{\kappa}
\def\l{\lambda}
\def\m{\mu}
\def\n{\nu}
\def\p{\psi}
\def\q{\partial}
\def\r{\rho}
\def\s{\sigma}
\def\t{\tau}
\def\u{\upsilon}
\def\v{\varphi}
\def\w{\omega}
\def\x{\xi}
\def\y{\eta}
\def\z{\zeta}
\def\D{\Delta}
\def\G{\Gamma}
\def\H{\Theta}
\def\L{\Lambda}
\def\F{\Phi}
\def\P{\Psi}
\def\S{\Sigma}

\def\o{\over}
\def\beq{\begin{eqnarray}}
\def\eeq{\end{eqnarray}}
\newcommand{\gsim}{ \mathop{}_{\textstyle \sim}^{\textstyle >} }
\newcommand{\lsim}{ \mathop{}_{\textstyle \sim}^{\textstyle <} }
\newcommand{\vev}[1]{ \left\langle {#1} \right\rangle }
\newcommand{\bra}[1]{ \langle {#1} | }
\newcommand{\ket}[1]{ | {#1} \rangle }
\newcommand{\EV}{ {\rm eV} }
\newcommand{\KEV}{ {\rm keV} }
\newcommand{\MEV}{ {\rm MeV} }
\newcommand{\GEV}{ {\rm GeV} }
\newcommand{\TEV}{ {\rm TeV} }
\def\diag{\mathop{\rm diag}\nolimits}
\def\Spin{\mathop{\rm Spin}}
\def\SO{\mathop{\rm SO}}
\def\O{\mathop{\rm O}}
\def\SU{\mathop{\rm SU}}
\def\U{\mathop{\rm U}}
\def\Sp{\mathop{\rm Sp}}
\def\SL{\mathop{\rm SL}}
\def\tr{\mathop{\rm tr}}

\def\IJMP{Int.~J.~Mod.~Phys. }
\def\MPL{Mod.~Phys.~Lett. }
\def\NP{Nucl.~Phys. }
\def\PL{Phys.~Lett. }
\def\PR{Phys.~Rev. }
\def\PRL{Phys.~Rev.~Lett. }
\def\PTP{Prog.~Theor.~Phys. }
\def\ZP{Z.~Phys. }

\newcommand{\beqr}{\begin{array}}  
\newcommand {\eeqr}{\end{array}}
\newcommand{\la}{\left\langle}  
\newcommand{\ra}{\right\rangle}
\newcommand{\non}{\nonumber}  
\newcommand{\ds}{\displaystyle}
\newcommand{\red}{\textcolor{red}}
\def\ubl{U(1)$_{\rm B-L}$}
\def\REF#1{(\ref{#1})}
\def\lrf#1#2{ \left(\frac{#1}{#2}\right)}
\def\lrfp#1#2#3{ \left(\frac{#1}{#2} \right)^{#3}}
\def\OG#1{ {\cal O}(#1){\rm\,GeV}}

\def\J{{\cal J}}
\def\aa{{\dot \a}}
\def\bb{{\dot \b}}
\def\ss{{\bar \s}}
\def\hh{{\bar \h}}
\def\Z{{\cal Z}}
\def\B{{\cal B}}

\newcommand{\Slash}[1]{{\ooalign{\hfil/\hfil\crcr$#1$}}}


\baselineskip 0.7cm

\begin{titlepage}

\begin{flushright}
\end{flushright}

\vskip 1.35cm
\begin{center}
{\large \bf
Perturbative $c$-theorem in $d$-dimensions}
\vskip 1.2cm
Kazuya Yonekura$~$ 
\vskip 0.4cm

{\it  School of Natural Sciences, Institute for Advanced Study, \\ Princeton, NJ 08540, USA\\
}

\vskip 1.5cm

\abstract{ 
We study perturbative behavior of free energies on a $d$-dimensional sphere $S^d$ for theories with marginal interactions.
The free energies are interpreted as the ``dilaton effective action''
with the dilaton having a nontrivial background vacuum expectation value. 
We compute the dependence of the free energies on
the radius of the sphere by using dimensional regularization.
It is shown that the first (second) derivative of the free energies in odd (even) dimensions with respect to the radius of the sphere
are proportional to 
the square of the beta functions of coupling constants. The result is consistent with the $c$, $F$ and $a$-theorems in two, three, four and six dimensions.
The result is also used to rule out a large class of scale invariant theories which are not conformally invariant.

}
\end{center}
\end{titlepage}

\setcounter{page}{2}

\section{Introduction}\label{sec:1}

In two dimensional quantum field theory, two elegant theorems are known.
Zamolodchikov showed~\cite{Zamolodchikov:1986gt} that there exists a function $c(r)$ of a length scale $r$ which monotonically decreases
as $r$ is increased, and becomes constant only on conformal fixed points. Roughly speaking, this result indicates that
a number of ``degrees of freedom'' monotonically decreases along renormalization group (RG) flows. This is the famous Zamolodchikov's $c$-theorem.
Then, Polchinski proved~\cite{Polchinski:1987dy} that all scale invariant theories (with discrete spectrum of scaling dimensions) are also conformally invariant.
The result of Ref.~\cite{Zamolodchikov:1986gt} played a crucial role in the proof of Ref.~\cite{Polchinski:1987dy}.

There have also been significant developments in the study of monotonically decreasing quantities in higher dimensions.
In even dimensional CFT, the trace of the energy-momentum tensor has anomaly when the CFT is coupled to external gravitational background.
It is given as~\cite{Deser:1993yx}
\beq
T^\m_\m=(-1)^{\frac{d}{2}+1} a E_d+\cdots, \label{eq:eventraceanomaly}
\eeq
where $E_{d} \propto \e_{\m_1\m_2 \cdots \m_{d-1}\m_{d}} \e^{\n_1\n_2 \cdots \n_{d-1}\n_{d}}R^{\m_1\m_2}_{~~~~\n_1\n_2}\cdots 
 R^{\m_{d-1}\m_{d}}_{~~~~~~~\n_{d-1}\n_{d}}$ is the Euler density, and the dots indicate terms which vanish in conformally flat background.
 In two dimensional CFTs, the coefficient of the Euler density $E_{d}$ in the trace anomaly (written as $a$ in Eq.~(\ref{eq:eventraceanomaly}))
 coincides with the Zamolodchikov's $c$ function.
 In general even dimensional field theories, 
 it was conjectured~\cite{Cardy:1988cwa} that $a$ decreases along RG flows in theories which interpolate UV and IR CFTs,
 that is, $a_{\rm IR}<a_{\rm UV}$. 

The quantity $a$ may be extracted in the following way. Let us put a CFT on a $d$-dimensional sphere $S^d$ with radius $r$ and consider
the partition function on the sphere,
\beq
Z=\int [D \varphi] e^{-S},
\eeq
where $\varphi$ denotes dynamical fields of the theory, $S$ is the action on the sphere, and $\int [D \varphi]$ is the path integral.
Using the fact that the change of the radius $r$ as $r \to e^{\s} r$ for a constant $\s$ is equivalent
to the Weyl rescaling of the metric as  $g_{\m\n} \to e^{2\s}g_{\m\n}$, the free energy $F=-\log Z$ satisfies
\beq
\frac{d F}{d \log r}=-\vev{ \int d^d \sqrt{g} T^\m_\m} \propto (-1)^{\frac{d}{2}} a,
\eeq
where we have used the fact that the metric of the sphere is conformally flat in projective coordinates, $ds^2=(2r^2/(x^2+r^2))^2dx^2$,
and hence the terms denoted by the dots in Eq.~(\ref{eq:eventraceanomaly}) do not 
contribute. Therefore, the above conjecture may be interpreted as the conjecture 
that the function $(-1)^{\frac{d}{2}} d F / d \log r$ decreases as $r$ is increased.

In odd dimensions, there is a similar conjecture about the free energy $F$~\cite{Jafferis:2011zi,Klebanov:2011gs}. 
In this case, it is $(-1)^{\frac{d+1}{2}} F$ that is conjectured to decrease.
Therefore, in both even and odd dimensions, the free energy on the sphere, $F=-\log Z$, plays an important role in
the study of monotonically decreasing quantities.

Recently, a proof that $a$ satisfies $a_{\rm IR}<a_{\rm UV}$ was given in four dimensions~\cite{Komargodski:2011vj}
and further discussed in Refs.~\cite{Komargodski:2011xv,Luty:2012ww}.
Also, a monotonically decreasing quantity was constructed in three dimensions~\cite{Casini:2012ei} which
coincides with $F$ in CFT~\cite{Casini:2011kv}.
Completely different methods were used in the proofs in two~\cite{Zamolodchikov:1986gt}, three~\cite{Casini:2012ei} and four~\cite{Komargodski:2011vj}
dimensions.
There is still no proof in general space-time dimensions, 
although holography suggests the existence of a monotonically decreasing function in arbitrary 
dimensions~\cite{Freedman:1999gp,Myers:2010xs,Myers:2010tj}.
See Refs.~\cite{Elvang:2012st,Elvang:2012yc,Bhattacharyya:2012tc} for recent works in six and higher dimensions.

Progress has also been made regarding the equivalence of scale and conformal invariance in four dimensions.
A proof of the equivalence was given in perturbation theory~\cite{Luty:2012ww,Fortin:2012hn} 
(see also Refs.~\cite{Callan:1970ze,Polchinski:1987dy,Dorigoni:2009ra,Antoniadis:2011gn,Zheng:2011bp}).\footnote{
Although the existence of explicit counterexamples are discussed~\cite{Fortin:2011ks,Fortin:2011sz,Fortin:2012ic,Fortin:2012cq},
they are argued to be conformally invariant~\cite{Nakayama:2011tk,Nakayama:2012nd,Fortin:2012hn} 
based on the results of Refs.~\cite{Jack:1990eb,Osborn:1991gm}.
}
Ref.~\cite{Luty:2012ww} also gave a non-perturbative argument in favor of the equivalence.
In that proof, the existence of a monotonically decreasing quantity $a$ (or more precisely the dilaton forward scattering amplitude) is essential.  
This is similar to the proof of the equivalence in two dimensions~\cite{Zamolodchikov:1986gt,Polchinski:1987dy}.

However, much less is known in other dimensions.\footnote{
There exist free field theory counterexamples in $d >4$~\cite{ElShowk:2011gz,Jackiw:2011vz}.
But there is no local current operator for scaling symmetry and only the charge is well-defined in those theories.
There still remains a possibility that every scale invariant theory with a scaling current is conformally invariant.}
In particular, there are many perturbative field theories
in three dimensions, and there is a possibility that 
 some of them could be scale invariant without conformal invariance by the same mechanism discussed in 
Refs.~\cite{Fortin:2011ks,Fortin:2011sz,Fortin:2012ic,Fortin:2012cq}.

As discussed above, one of the ways to generalize the two dimensional theorems to arbitrary dimensions may be to use the free energy on the sphere.
The flow of the free energy was studied when a CFT is deformed by adding slightly marginal operators ${\cal O}$ 
to the Lagrangian~\cite{Cardy:1988cwa,Klebanov:2011gs}. The operators were assumed to have scaling dimensions
$d-y$ with $y \ll 1$.

In this paper, we study the free energy for general weakly interacting field theories with marginal interactions.
A list of such theories is given in Table.~\ref{table:1}. 
We show that $(-1)^{\frac{d+1}{2}} F$ (in odd dimensions) or $(-1)^{\frac{d}{2}} d F / d \log r$ (in even dimensions) decreases monotonically
in these theories. Furthermore, following Ref.~\cite{Luty:2012ww}, we argue that scale invariance is equivalent to conformal invariance in these theories.

The rest of the paper is organized as follows. In section~\ref{sec:2} we give a relation between the free energy on the sphere
and the ``dilaton effective action'' which was used in the proof of the $a$-theorem in four dimensions~\cite{Komargodski:2011vj,Komargodski:2011xv,Luty:2012ww}.
It enables us to compute perturbative flows of the free energy by using the method of Refs.~\cite{Komargodski:2011xv,Luty:2012ww}.
We obtain the dilaton effective action in dimensional regularization.
In section~\ref{sec:3}, we compute the flow of the free energy using the dilaton effective action. We check our result in two, three, four and six dimensions.
Using the result, we argue the equivalence of scale and conformal invariance.
Section~\ref{sec:4} is devoted to conclusions.

\begin{table}[]
\begin{center}
\begin{tabular}{|c|c|}
\hline
Dimensions & Lagrangians   \\
\hline
 $d=2$ & $G_b(\phi)(\q_\m\phi)^2 +G_f(\phi)\psi \Slash{\q}\psi+H(\phi)\psi^4$\\
\hline
$d=3$ & $ (A \q A+\frac{2}{3}A^3)+(D_\m \phi)^2+\psi \Slash{D}\psi+\phi^6+\phi^2 \psi^2 $ \\
\hline 
$d=4$ & $(F_{\m\n})^2+(D_\m \phi)^2+\psi \Slash{D}\psi+\phi^4+ \phi \psi^2$  \\
\hline
$d=6$ &  $ (\q_\mu \phi)^2+\phi^3$     \\
\hline
\end{tabular}
\caption{Schematic forms of the Lagrangians of perturbative theories with marginal interactions. 
The fields $\phi$ are bosons, $\psi$ are fermions, $A$ are gauge bosons, $F$ are gauge field strengths, and
possible indices specifying these fields are suppressed. $G_b(\phi)$, $G_f(\phi)$ and $H(\phi)$ are arbitrary functions of scalar fields $\phi$.
All the coupling constants are dimensionless in these Lagrangians.}
\label{table:1}
\end{center}
\end{table}

\section{Dilaton effective action}\label{sec:2}
We define the free energy of a theory on a $d$-dimensional sphere as a dilaton effective action in the following way.
We first consider the partition function as a functional of a background metric $\hat{g}_{\m\n}$. (The hat is used on the metric
following the notation of Refs.~\cite{Komargodski:2011vj,Komargodski:2011xv,Luty:2012ww}.)
It is given as
\beq
Z &=& \int [D\varphi] \exp\left(-S[\varphi,\hat{g}_{\mu\nu}]-S_{\rm c.t.}[\hat{g}_{\m\n}] \right) \nonumber \\
&=& Z_0  \exp\left(-S_{\rm eff,0}[\hat{g}_{\mu\nu}]-S_{\rm c.t.}[\hat{g}_{\m\n}] \right),
\eeq
where $\varphi$ denotes dynamical fields of the theory, and $S[\varphi,\hat{g}_{\mu\nu}]$ is the action of the fields $\varphi$ coupled to the metric $\hat{g}_{\m\n}$.
The factor $Z_0$ is the contribution to the partition function which does not depend on the background metric,
and $S_{\rm eff,0}$ is the (bare) effective action of the metric obtained as a result of the path integral.
The counterterm $S_{\rm c.t.}$ is taken so that the functional
\beq
S_{\rm eff}[\hat{g}_{\m\n}]=S_{\rm eff,0}[\hat{g}_{\mu\nu}]+S_{\rm c.t.}[\hat{g}_{\m\n}] ,
\eeq
becomes finite. We will impose further condition on the counterterms $S_{\rm c.t.}$ later.

We introduce a dilaton field $\t$ and a new metric $g_{\m\n}$ as $\hat{g}_{\m\n}=e^{-2\tau} g_{\mu\nu}$.
Then the dilaton effective action is defined as
\beq
S[\tau, g_{\m\n}]=S_{\rm eff}[\hat{g}_{\m\n}=e^{-2\tau} g_{\mu\nu}].
\eeq
This definition of the dilaton effective action is emphasized in Ref.~\cite{Luty:2012ww}. 
When $g_{\mu\nu}=\y_{\m\n}$, it gives the dilaton effective action in flat space, and 
this definition makes clear the invariance of the dilation effective action
under conformal transformations. This is because conformal transformations are just 
the subgroup of the diffeomorphism of the original metric $\hat{g}_{\m\n}$
which preserves the form $d\hat{s}^2=e^{-2\t}dx^2$.

The metric of the sphere with radius $r$ can be written using the projective coordinates as
$d\hat{s}^2=\hat{g}_{\mu\nu}dx^\m dx^\n=[2r^2/(x^2+r^2)]^2 dx^2$. However, we may also interpret this
as a flat metric $g_{\m\n}=\y_{\m\n}$ with a nontrivial background for the dilaton, $e^{-\t}=2r^2/(x^2+r^2)$.
Then, the free energy of the theory on the sphere, $F=-\log Z$, is given as 
\beq
F(r)=-\log Z_0+S\left[ e^{-\tau}=\frac{2r^2}{x^2+r^2}, \y_{\m\n} \right]. \label{eq:deffreeenergy}
\eeq
By this interpretation, we can use the results of Refs.~\cite{Komargodski:2011xv,Luty:2012ww} for the dilaton effective action
to compute the free energy on the sphere.

It is clear that the dependence of $F(r)$ on the radius of the sphere $r$ should be contained in the second term
of Eq.~(\ref{eq:deffreeenergy}). 
In this paper we attempt to calculate only the derivatives of
$F(r)$ with respect to $r$. Then we may neglect the term $\log Z_0$, and focus on the dilaton effective action.

The above definition still has an ambiguity regarding the choice of the counterterms in $S_{\rm c.t.}$.
Although the divergent part of $S_{\rm c.t.}$ is determined uniquely so that it makes the metric effective action $S_{\rm eff}$ finite,
the finite part of $S_{\rm c.t.}$ is not fixed. We impose the following requirement on the finite part.
In this paper we only consider massless theories which do not contain dimensionful parameters (see Table.~\ref{table:1}).
Furthermore, we always use dimensional regularization as a regularization method. 
Then, by using mass-independent renormalization scheme (such as minimal subtraction), counterterms 
which contain dimensionful coefficients are not necessary (see e.g. Ref.~\cite{Weinberg:1996kr}). That is, we can set all the counterterms to zero 
aside from counterterms with dimensionless coefficients which are schematically given as
\beq
S_{\rm c.t.}[\hat{g}_{\m\n}] \sim \int d^dx \sqrt{\hat{g}} (R_{\m\n\r\s}(\hat{g}))^{\frac{d}{2}}, \label{eq:massindependentcounterterm}
\eeq
where $R_{\m\n\r\s}$ is the Riemann tensor and indices are contracted in arbitrary ways. 
Therefore, we only introduce counterterms of the form (\ref{eq:massindependentcounterterm}). 
This is our criterion for choosing the counterterms.

In the case of odd dimensions, terms like Eq.~(\ref{eq:massindependentcounterterm}) do not exist and hence
we need no counterterms at all. Therefore $F(r)$ is uniquely determined by our criterion. 
In even dimensions, finite counterterms of the form (\ref{eq:massindependentcounterterm}) are allowed,\footnote{The finite counterterms
are in fact necessary in order for the effective action $S_{\rm eff}(\hat{g}_{\m\n})$ to be RG invariant. Even if they are set to zero at some RG scale,
they are generated along RG flows.}
and hence the ambiguity in defining $F(r)$ remains.
However, one can see that the contributions coming from these finite counterterms disappear if we take the derivative of the free energy, 
$dF/dr$.\footnote{Strictly speaking, these contributions are not precisely zero in dimensional regularization. 
They are suppressed by $\e$, where the space-time dimensions is given by $d=({\rm integer})-2\e$.
Then, the contributions of the finite part of the counterterms become zero when we take $\e \to 0$, but the contributions from divergent part of the counterters are
important. 
} As discussed in the introduction, the important quantity in even dimensions is $dF/dr$ rather than $F$ itself, and hence
the remaining ambiguity in choosing the counterterms does not matter.

The above requirement on the counterterms is a little technical. More physical requirement may be that the free energy $F$ (in odd dimensions)
or its derivative $dF/dr$ (in even dimensions) becomes constant on UV/IR fixed points. This physical requirement will be satisfied by the above
choice of the counterterms.

Now let us study the dilaton effective action $S[\tau]$ in flat (Euclidean) space-time.
The most important part of $S[\tau]$ in perturbation theory has been given in Refs.~\cite{Komargodski:2011xv,Luty:2012ww}.
We use dimensional regularization where we work in
$d=d_0-2\e$ dimensions with $d_0 $ an integer.

We expand the action as
\beq
S[\varphi, \hat{g}_{\m\n}= e^{-2\tau}\y] &=& S[\varphi, \y_{\m\n}] +\int d^d x \t T^{\m}_{\m}+O(\t^2), \label{eq:mattertau}
\eeq
where $T^{\m\n}$ is the energy-momentum tensor.
The linear term in $\tau$ is proportional to the trace anomaly. 
We assume that interaction terms are present in the Lagrangian as $\lambda_0^i {\cal O}_i $,
where $\lambda^i_0$ are bare couplings and ${\cal O}_i$ are bare operators.
For example, in a four-dimensional scalar $\phi^4$ theory, we may define ${\cal O}=\frac{1}{4!} \phi^4$.
Then, if the energy-momentum tensor is improved appropriately, the trace anomaly may be given as 
\beq
T^\m_\m=-\sum_i \B^i [{\cal O}_i], \label{eq:tracetmtensor}
\eeq
where $[{\cal O}]_i$ are the renormalized operators corresponding to the bare operator ${\cal O}_i$, and $\B^i$ are the beta functions of the 
renormalized coupling constants $\l^i$.
In cases where there are many flavors of matter fields, there are ambiguities in the definition of  usual beta functions $\b^i$,
while the beta functions $\B^i$ appearing in Eq.~(\ref{eq:tracetmtensor}) is unambiguous~\cite{Jack:1990eb,Osborn:1991gm}. 
Following the notation of Refs.~\cite{Jack:1990eb,Osborn:1991gm}, we denote this unambiguous beta functions as $\B^i$ rather than $\b^i$.

The improvement of the energy-momentum tensor is related to the term $R \phi^2$ in the Lagrangian, where $R$ is the Ricci scalar and $\phi$
are scalar fields of the theory.
For a moment, let us assume that this term is chosen so that Eq.~(\ref{eq:tracetmtensor}) holds.
We will revisit this point at the end of this section.

The higher order terms of $\t$ in Eq.~(\ref{eq:mattertau}) are accompanied by additional powers of $\e$ or
the coupling constants~\cite{Luty:2012ww}.  
The reason is the following.
In theories with only dimensionless parameters which are listed in Table.~\ref{table:1}, 
the dilaton appears in the combination $\e \t$ in the bare Lagrangian after performing
appropriate Weyl rescaling of matter fields.
For example, in the case of a four dimensional $\phi^4$ theory, the bare Lagrangian
of the theory is given as \footnote{Here we pretend as if the term $R \phi^2$ is chosen as the conformal coupling of a
free scalar, $\frac{(d-2)}{8(d-1)}R \phi^2$. This is not correct at higher orders of perturbation theory~\cite{Collins:1976vm}, but the corrections
occur at sufficiently higher orders so that the following discussion is not violated. }

\beq
\sqrt{\hat{g}} \left( \frac{1}{2}\hat{g}^{\m\n}\q_\m \hat{\phi}\q_\n \hat{\phi}+\frac{d-2}{8(d-1)}R(\hat{g}) \hat{\phi}^2 +\frac{\l_0}{4!}\hat{\phi}^4 \right)=
 \left( \y^{\m\n} \q_\m \phi\q_\n \phi +e^{-2\e \t} \frac{\l_0}{4!}\phi^4 \right)
\eeq
where $\hat{\phi}$ is the bare field, $\hat{g}_{\m\n}=e^{-2\t}\y_{\m\n}$ and $\phi=e^{-(d-2)\t/2}\hat{\phi}$.
Loop calculations give divergences which may cancel the factor $\e$ in $\e \t$. However, whenever $\e$ is cancelled,
there is always an additional loop suppression factor $L$ (e.g., $L=\l/16\pi^2 $ in the $\phi^4$ theory).
Therefore, $\t$ appears only in the combination $\e \t$ or $L \t$.

Then, neglecting the higher order terms, the leading order term in the dilaton effective action is given by
\beq
S_{\rm eff,0}[\hat{g}_{\mu\nu}=e^{-2\tau}\y_{\m\n}]=-\frac{1}{2}\int d^d x d^d y \t(x) \t(y) \sum_{i,j}\B^i\B^j\vev{ [{\cal O}_i(x)] [{\cal O}_i(y)] }+\cdots
\eeq
where dots denote higher order terms in $\e$ or loop factors.

At the leading order of perturbation theory, correlation functions of $[{\cal O}_i]$ are given as
\beq
\vev{ [{\cal O}_i(x)] [{\cal O}_i(y)] }=\frac{\mu^{2(d-d_i)}c_i \d_{ij}}{|x-y|^{2d_i}}, \label{eq:operatorcorrelation}
\eeq
where $c_i$ are dimensionless constants (e.g., $c=\frac{1}{4!}(\G(d/2-1)/4\pi^{d/2})^4$ for ${\cal O}=\frac{1}{4!} \phi^4$
), 
$\m$ is the unit of mass of dimensional regularization
(or in other words the RG scale), and $d_i$ is the classical scaling dimension of ${\cal O}_i$.
Although we are considering only marginal interactions, $d_i$ differs from $d$ by order $\e$ (e.g., $d_i=2(d-2)=4-4\e$ for ${\cal O}=\frac{1}{4!} \phi^4$). 
The operators $[{\cal O}_i]$ are assumed to be normalized so that 
$\vev{ [{\cal O}_i(x)] [{\cal O}_i(y)] } $ is proportional to $\d_{ij}$ at the leading order.
The constants $c_i$ are ensured to be positive by reflection positivity. Then the dilaton effective action becomes
\beq
S_{\rm eff,0}[\hat{g}_{\mu\nu}=e^{2\tau}\y_{\m\n}] &=&-\frac{1}{2}\int d^d x d^d y \t(x) \t(y) \sum_{i}\frac{\mu^{2(d-d_i)}c_i \B_i^2}{|x-y|^{2d_i}}+\cdots \nonumber \\
&=&-
\int \frac{d^d k}{(2\pi)^d} |\tilde{\t}(k) |^2 \sum_{i}\frac{\pi^{d \o 2} 2^{d-2d_i} \Gamma(d/2-d_i)}{2\Gamma(d_i)}c_i \B_i^2\mu^{2(d-d_i)}k^{2d_i-d}+\cdots 
\nonumber \\ \label{eq:baredilatonaction}
\eeq
where we have Fourier-transformed the dilaton as $\tilde{\t}(k)=\int d^d x e^{-ikx}\t(x)$. 

In odd dimensions, the above dilaton effective action is finite in the limit $\e \to 0$. This is consistent with the fact that we need no counterterms 
in odd dimensions as discussed above. In even dimensions, there is a divergence coming from the factor $ \Gamma(d/2-d_i)$
and we have to renormalize it. The counterterm should be local and is given as
\beq
S_{\rm c.t.}[\hat{g}_{\m\n}=e^{-2\tau}\y_{\m\n}] &=&
\int \frac{d^d k}{(2\pi)^d} |\tilde{\t}(k) |^2 \left(a_0+ \sum_{i}\frac{\pi^{d \o 2} 2^{d-2d_i} \Gamma(d/2-d_i)}{2\Gamma(d_i)} c_i \B_i^2\right)\mu^{d-d_0}k^{d_0}+\cdots
\nonumber \\
&=&\int d^d x \t(x)(-\q^2)^{d_0 \o 2}\t(x)  \left(a_0+ \sum_{i}\frac{\pi^{d \o 2} 2^{d-2d_i} \Gamma(d/2-d_i)}{2\Gamma(d_i)} c_i \B_i^2\right)\mu^{d-d_0}+\cdots
\nonumber \\ \label{eq:dilatonconter}
\eeq
where $a_0$ is a constant which is finite in the limit $\e \to 0$. This counterterm makes the dilaton effective action finite.

Although it is not immediately evident whether the counterterm (\ref{eq:dilatonconter}) can be obtained from counterterms for the metric 
$S_{\rm c.t.}[\hat{g}_{\m\n}] $ by replacing the metric as $\hat{g}_{\m\n} \to e^{-2\tau}\y_{\m\n}$, 
it is known to be possible~\cite{Elvang:2012st,Elvang:2012yc}. 
It may be instructive to see it explicitly
in the simplest case where the space-time dimension is $d=2-2\e$.
There is only one candidate for the counterterm which is given by
\beq
S_{\rm c.t.}[\hat{g}_{\m\n}] \propto \int d^{d} x \sqrt{\hat{g}}R(\hat{g}).
\eeq
Then, the dilaton counterterm is obtained as 
\beq
S_{\rm c.t.}[\hat{g}_{\m\n}=e^{-2\tau}g_{\m\n}]  & \propto & \int d^{d} x \sqrt{g}
e^{2\e\t}\left[R(g)- 2\e(1-2\e) (\nabla \t)^2 \right]\nonumber \\
&=& \int d^{d} x \sqrt{g}\left[R(g)+2\e \left(\t R(g) -(\nabla \t)^2 \right) +O(\e^2) \right]. \label{eq:twodimcounter}
\eeq
Thus, by taking $g_{\m\n} \to \y_{\m\n}$, the counterterm of the form $\frac{1}{\e} \t \q^2 \t$ (a single pole term in $\e$) is obtained from 
$\frac{1}{\e^2} \sqrt{\hat{g}}R(\hat{g})$ (a double pole term in $\e$). 

One should also notice that the finite term
in the dilaton counterterm $S_{\rm c.t.}[\hat{g}_{\m\n}=e^{-2\tau}\y_{\m\n}] $ actually comes from the divergent term in the metric counterterm 
$S_{\rm c.t.}[\hat{g}_{\m\n}]$. 
In fact, this is how the Wess-Zumino action for the dilaton~\cite{Schwimmer:2010za,Komargodski:2011vj} arises in dimensional regularization.
The integral of the Euler density
 $E_{d_0}$
 is a topological quantity in $d_0$-dimensions.
In the case of $d=d_0-2\e$ dimensional space-time, this topological property is broken by $\e$, and the change of the metric
$\hat{g}_{\m\n} = e^{-2\t}g_{\m\n}$ gives
\beq
\int d^d x \sqrt{\hat{g}}E_{d_0}(\hat{g}) = \int d^d x \sqrt{g}E_{d_0}(g)+2\e S_{\rm WZ}+O(\e^2), \label{eq:dilatonWZ}
\eeq 
where
\beq
S_{\rm WZ}=\int d^dx  \sqrt{g} \left( \t E_{d_0}(g)+\cdots \right)
\eeq
is the Wess-Zumino action for the dilaton. See Eq.~(\ref{eq:twodimcounter}) for the case of $d_0=2$.
In CFTs, we need a counterterm of the form $\int d^d x(a / \e)E_{d_0}$ to make the energy-momentum tensor
finite~\cite{Duff:1993wm}. This counterterm leads to the trace anomaly $T^\m_\m \sim a E_{d_0} $. One can see that the presence of this counterterm
gives the finite Wess-Zumino action $a S_{\rm WZ}$ for the dilaton by using Eq.~(\ref{eq:dilatonWZ}).

Let us return to the computation of the dilaton effective action. We have neglected higher order terms in $\e \t$ and loop factors.
We continue to neglect the higher order corrections of the loop factors. 
However, for our purposes it is important to recover the higher order terms of $\e \t$.
Actually, there are divergences when we compute the free energy by substituting $e^{-\t}=2r^2/(x^2+r^2)$.  
It turns out that terms containing extra powers of $\t=\log ((x^2+r^2)/2r^2)$, $\t^k $~($k=0,1,2,\cdots$), give additional divergences $\frac{1}{\e^k}$.
Therefore it is necessary to retain higher order terms in $\e \t$.

To recover the dependence on $\e \t$, we use the conformal invariance of the dilaton effective action. As discussed above,
the dilaton effective action should be conformally invariant since the conformal transformations are just the subgroup of the diffeomorphism 
of the original effective action for the metric. We can make Eqs.~(\ref{eq:baredilatonaction}) and (\ref{eq:dilatonconter}) conformally invariant
by replacing them as
\beq
\int d^d x \t(x)  \frac{1}{|x-y|^{2d_i}} \t(y)\to \int d^d x d^d y \left( \frac{e^{-(d-d_i)\t(x)}}{d-d_i} \right) \frac{1}{|x-y|^{2d_i}} \left( \frac{e^{-(d-d_i)\t(y)}}{d-d_i} \right),
\label{eq:replaceaction}
\eeq
and
\beq
\int d^d x \t(x) (-\q^2)^{d_0 \o 2} \t(y)\to \int d^d x  \left( \frac{e^{-(\frac{d-d_0}{2})\t(x)}}{(d-d_0)/2} \right) (-\q^2)^{d_0 \o 2} \left( \frac{e^{-(\frac{d-d_0}{2})\t(x)}}{(d-d_0)/2} \right).
\label{eq:replacecounterterm}
\eeq
By expanding in $\t$,
one can check that the linear terms in $\t$ are absent due to the properties of dimensional regularization. 
The quadratic terms in $\t$ just reproduce the original ones. Higher order terms in $\t$ are accompanied by appropriate powers of $d-d_i \propto \e$ as expected. 
The modified action in the right-hand-side of Eq.~(\ref{eq:replaceaction}) is conformally invariant since
the field $e^{-(d-d_i)\t}$ has the scaling dimension $(d-d_i)$. By performing Fourier transformation to momentum space, 
the right-hand-side of Eq.~(\ref{eq:replacecounterterm})
is just the same as that of Eq.~(\ref{eq:replaceaction}) by analytically continuing $d_i \to(d+ d_0)/2$
(up to a field-independent factor). Therefore it is also conformally invariant. See Refs.~\cite{Elvang:2012st,Elvang:2012yc} for the construnction
of this counterterm from $S_{\rm c.t.}[\hat{g}_{\m\n}]$.
By using the replacements (\ref{eq:replaceaction}) and (\ref{eq:replacecounterterm}) in Eqs.~(\ref{eq:baredilatonaction}) and (\ref{eq:dilatonconter})
respectively, we obtain the desired dilaton effective action with nonzero $\e$.

Before closing this section, let us discuss the term $R \phi^2$ in the matter action.
In general, the existence of this term gives an additional ambiguity in the definition of the free energy
because we can introduce new parameters $\xi$ as $\xi R \phi^2$.\footnote{
In the case of the six dimensional $\phi^3$ theories, there can also exist linear terms of $\phi$ given by $ \eta R\nabla^2 \phi$ and $\zeta R^2 \phi$.
Just for simplicity, we consider the case $d \leq 4$ in the following discussion.}
This is related to the ambiguity of the energy-momentum tensor $T_{\m\n}$, since
the addition of the term $\xi R \phi^2$ changes $T_{\m\n}=\frac{2}{\sqrt{g}}\frac{\d S}{\d g^{\m\n}}$ 
by improvement term of the form $(\q^\m \q^\n-\y^{\m\n}\q^2)\xi \phi^2$.

We assume the existence of the renormalization scheme for the energy-momentum tensor $T_{\m\n}$ such that
\begin{enumerate}
\item $T_{\m\n}$ is RG invariant, i.e., $\m\frac{\q}{\q\m} T_{\m\n}=0$. In other words, there is no operator mixing with 
$(\q^\m \q^\n-\y^{\m\n}\q^2)\phi^2$.
\item The trace anomaly (with nontrivial metric) is given by
\beq
T^{\m}_\m=-\B^i[{\cal O}_i]'+{\rm purely~metric~terms}, \label{eq:curvedtrace}
\eeq
where $[{\cal O}_i]'$ are operators which coincide with $[{\cal O}_i]$ in the flat space limit.
In particular, $T^\m_\m$ becomes metric dependent c-number when $\B_i=0$. 
\end{enumerate}
(Actually, in $d \leq 4$ dimensions it is enough that Eq.~(\ref{eq:curvedtrace}) is satisfied in flat space, since
the flat space trace anomaly combined with Wess-Zumino consistency condition for Weyl transformations leads to Eq.~(\ref{eq:curvedtrace})
in nontrivial metric background~\cite{Osborn:1991gm}.)
We always couple the metric to the energy-momentum tensor satisfying the above assumptions.

The existence of the energy-momentum tensor satisfying the above assumptions can be explicitly proved for some theories.
For example, a proof was given in Ref.~\cite{Collins:1976vm} for the case of $\phi^4$ theory in four dimensions.
A large class of four dimensional renormalizable supersymmetric theories also satisfies the assumptions. 
In this case there is a Ferrara-Zumino supercurrent multiplet~\cite{Ferrara:1974pz}, ${\cal J}^{\rm FZ}_{\a\dot{\a}}$
(if the theory does not contain FI-terms~\cite{Komargodski:2009pc}.
See also Refs.~\cite{Komargodski:2010rb,Dumitrescu:2011iu} for recent comprehensive discussions on supercurrents.)
The Ferrara-Zumino supercurrent multiplet can mix with other operators only as 
${\cal J}^{\rm FZ}_{\a\dot{\a}} \to {\cal J}^{\rm FZ}_{\a\dot{\a}}+[D_\a , \bar{D}_{\dot{\a}}](\Phi+\Phi^\dagger)$,
where $\Phi$ is a chiral superfield, i.e, $\bar{D}_{\dot{\a}}\Phi=0$, and $\Phi$ has mass dimension two.
If there is no candidate for $\Phi$ which is invariant under any global or local symmetries, 
the multiplet ${\cal J}^{\rm FZ}_{\a\dot{\a}}$ cannot mix with any other operators. Then the energy-momentum tensor 
contained in ${\cal J}^{\rm FZ}_{\a\dot{\a}}$ satisfies the first assumption.
It also satisfies the second assumption~\cite{Clark:1978jx,Piguet:1981mu,Piguet:1981mw}.\footnote{In higher orders of perturbation theory,
there is a notorious problem called the anomaly puzzle~\cite{Grisaru:1978vx}. 
See Refs.~\cite{Grisaru:1985ik,Shifman:1986zi,ArkaniHamed:1997mj,Huang:2010tn,Yonekura:2010mc,Yonekura:2012uk} and
references therein for discussions on this problem.
} 
Thus a large class of supersymmetric theories has 
the energy-momentum tensor satisfying the above assumptions. The situation is similar in three dimensions as long as the Ferrara-Zumino multiplet exists.

If we introduce additional parameters $\x$, it is possible to construct an RG invariant energy-momentum tensor. 
Let ${\cal O}^{\phi^2}_a$ denote the set of operators $\phi^2$, where $a$ is the label specifying the operators. 
If the energy-momentum tensor satisfies Eq.~(\ref{eq:curvedtrace}), the operator mixing is in general given by (see e.g., Ref.~\cite{Osborn:1991gm})
\beq
\m \frac{\q}{\q \mu} 
\left(
\begin{array}{c}
~T_{\m\n} \\
~[{\cal O}_i] \\
~[{\cal O}^{\phi^2}_a]
\end{array}
\right)
=
\left(
\begin{array}{ccc}
0 & 0 & -\B^k \d^b_k(\y^{\m\n}\q^2-\q^\m \q^\n)/(d-1) \\
0 & -\q_i \B^j & \d^b_i \q^2 \\
0 & 0 & \g^b_a
\end{array}
\right)
\left(
\begin{array}{c}
~T_{\m\n} \\
~[{\cal O}_j] \\
~[{\cal O}^{\phi^2}_b]\end{array}
\right),
\eeq
where $\d^a_i$ and $\g^b_a$ are some anomalous dimension matrices, and $\q_i \B^j$ is the derivative of $ \B^j$ 
with respect to the coupling $\l^i$. 
We introduce new parameters $\xi^a_i$ which are defined to satisfy the RG equation
\beq
\mu \frac{\q}{\q \mu}\xi^a_i  + \g^a_b \xi^b_i+\q_i \B^j \xi^a_j+\d^a_i=0. \label{eq:additionalparameter}
\eeq
Then, we define the new operators as
\beq
[{\cal O}^{\rm (new)}_i] &=& [{\cal O}_i]+\xi^a_i \q^2 [{\cal O}^{\phi^2}_a], \\
T^{\rm (new)}_{\m\n} &=& T_{\m\n}-\frac{1}{d-1}\B^i \xi^a_i(\y^{\m\n}\q^2-\q^\m \q^\n)  [{\cal O}^{\phi^2}_a]. \label{eq:newtensor} 
\eeq
This new energy-momentum tensor is invariant under RG and has a trace $-\B^i [{\cal O}^{\rm (new)}_i] $. 
If we can find $\xi^a_i$ as a function of the coupling constants
$\l^i$ with a well-defined perturbative expansion, the difference between $T_{\m\n}$ and $T^{\rm (new)}_{\m\n}$ is just a change of
renormalization prescription and $T^{\rm (new)}_{\m\n}$ is the desired energy-momentum tensor.
This was indeed shown to be possible in $\phi^4$ theory~\cite{Collins:1976vm}.
Even if $\xi^a_i$ is not a function of $\l^i$, our computation of the free energy is valid as long as we can find a solution to 
Eq.~(\ref{eq:additionalparameter}) such that $\xi^a_i$ remain small along RG flows.
One can check that $\d^a_i$ vanish at the one loop level, and hence nonzero $\xi^a_i$ are generated only at higher orders of perturbation theory.

\section{Perturbative free energy and $c$-theorems}\label{sec:3}
\subsection{Free energy in $d$-dimensions}

Let us summarize the result for the leading term of the dilaton effective action obtained in the previous section. 
The unrenormalized effective action is given by
\beq
S_{\rm eff,0}[\hat{g}_{\mu\nu}=e^{-2\tau}\y_{\m\n}] =
-\frac{1}{2 }\sum_{i} c_i \B_i^2 I_{d_i}  ,
\eeq
and the counterterm (for $d_0$=even) is given by
\beq
S_{\rm c.t.}[\hat{g}_{\m\n}=e^{-2\tau}\y_{\m\n}]=\left(a_0+\sum_{i} \frac{\pi^{d \o 2} 2^{d-2d_i} \Gamma(d/2-d_i)}{2\Gamma(d_i)}c_i \B_i^2 \right)  J,
\eeq
where $I_{d_i}$ and $J$ are defined as 
\beq
I_{d_i} &=& \mu^{2(d-d_i)}\int d^d x d^d y \left( \frac{e^{-(d-d_i)\t(x)}}{d-d_i} \right) \frac{1}{|x-y|^{2d_i}} \left( \frac{e^{-(d-d_i)\t(y)}}{d-d_i} \right), \\
J &=&\mu^{(d-d_0)} \int d^d x \left( \frac{e^{-(\frac{d-d_0}{2})\t(x)}}{(d-d_0)/2} \right) (-\q^2)^{d_0 \o 2} \left( \frac{e^{-(\frac{d-d_0}{2})\t(y)}}{(d-d_0)/2} \right).
\eeq
In this section we evaluate the explicit values of $I_{d_i}$ and $J$ when we substitute the dilation vacuum expectation value
$e^{-\t}=2r^2/(x^2+r^2)$.

The computation of $I_{d_i}$ is the same as in Refs.~\cite{Cardy:1988cwa,Klebanov:2011gs}.
First we rewrite $I_{d_i}$ as
\beq
I_{d_i} &=&\frac{\mu^{2(d-d_i)}}{(d-d_i)^2}\int d^d x \sqrt{\hat{g}(x)}  d^d y \sqrt{\hat{g}(y)}  \frac{1}{s(x,y)^{2d_i}}, \\
s(x,y) &=&  \frac{2r^2}{(x^2+r^2)^{1 \o 2}(y^2+r^2)^{1 \o 2}} |x-y|.
\eeq
where $\hat{g}_{\m\n}=e^{-2\t} \y_{\m\n}$ is the metric of the sphere, and $s(x,y)$ is the ``chordal distance'' between points $x$ and $y$
if the sphere is embedded in a flat Euclidean space. Then, by using the rotational invariance of the sphere, we may set $y=\infty$ to obtain
\beq
I_{d_i} &=&\frac{\mu^{2(d-d_i)}{\rm Vol.}(S^d)}{(d-d_i)^2} \int d^d x \sqrt{\hat{g}(x)}   \frac{1}{s(x,\infty)^{2d_i}} \nonumber \\
&=& \frac{\mu^{2(d-d_i)}}{(d-d_i)^2} \frac{2\pi^{\frac{d+1}{2}} r^{d}  }{\Gamma(\frac{d+1}{2}) } \int d^dx  \frac{(2r^2)^{d-2d_i}}{(x^2+r^2)^{(d-d_i)}} \nonumber \\
&=&(2\mu r)^{2(d-d_i)} \frac{  \pi^d   \G(\frac{d}{2})\G(\frac{d}{2}-d_i)    }{(d-d_i) \Gamma(d) \G(d-d_i+1) } ,
\eeq
where ${\rm Vol.}(S^d)$ is the volume of the sphere with radius $r$, and
in the process of the computation we have used some identities of the gamma function such as $\G(\frac{d}{2})\G(\frac{d+1}{2})=\pi^{1 \o 2}2^{1-d}\G(d)$.

It is also easy to compute $J$. By Fourier transforming to momentum space and looking at the expressions for $I_{d_i}$ and $J$ in momentum space, 
we find that $I_{d_i} $ and $J$ are related as
\beq
J &=& \lim_{d_i \to {d+d_0 \o 2}} \frac{\Gamma(d_i)}{\pi^{d \o 2} 2^{d-2d_i} \Gamma(d/2-d_i)} I_{d_i} \nonumber \\
&=& (2\mu r)^{(d-d_0)} \frac{ 2^{d_0 }  \pi^{d \o 2}   \G(\frac{d}{2}) \Gamma(\frac{d+d_0}{2})  }{(\frac{d-d_0}{2}) \Gamma(d) \G(\frac{d-d_0}{2}+1) }.
\eeq

By combining the above results, we finally get the free energy $F=-\log Z$ for odd dimensions or its derivative $d F/d \log r $ for even dimensions
as follows.\\
{\bf Odd dimensions}
\beq
F_{d_0={\rm odd}} &=& 
-\log Z_0-\frac{1}{2 }\sum_{i} c_i \B_i^2  (2\mu r)^{2(d-d_i)} \frac{  \pi^d   \G(\frac{d}{2})\G(\frac{d}{2}-d_i)    }{(d-d_i) \Gamma(d) \G(d-d_i+1) } 
+\cdots\nonumber \\
&=& ({\rm const.})+(-1)^{d_0-1 \o 2}   \frac{  2 \pi^{d_0+1}  }{ d_0 !   } \log(\mu r)\sum_{i} c_i \B_i^2 +\cdots.
\eeq
{\bf Even dimensions}
\beq
\frac{d F_{d_0={\rm even}}}{d \log r}
&=&(2\mu r)^{(d-d_0)} \frac{ 2^{d_0 }  \pi^{d \o 2}   \G(\frac{d}{2}) \Gamma(\frac{d+d_0}{2})  }{\Gamma(d) \G(\frac{d-d_0}{2}+1) }
\left(2a_0+\sum_{i} \frac{\pi^{d \o 2} 2^{d-2d_i} \Gamma(d/2-d_i)}{\Gamma(d_i)}c_i \B_i^2 \right) \nonumber \\
&&-\sum_{i} c_i \B_i^2  (2\mu r)^{2(d-d_i)} \frac{  \pi^d   \G(\frac{d}{2})\G(\frac{d}{2}-d_i)    }{\Gamma(d) \G(d-d_i+1) } +\cdots \nonumber \\
&=&({\rm const.})+(-1)^{\frac{d_0}{2}+1} \frac{4 \pi^{d_0}}{d_0 !} \log(\m r)  \sum_{i} c_i \B_i^2  +\cdots
\eeq
The dots denote higher order corrections in $\e$ or loop factors.

The constant terms in the above equations depend on $\log Z_0$ (in odd dimensions) or $a_0$ (in even dimensions) which we have not computed.
However, from the above result we can obtain the flows of $F$ or $d F/d \log r$ as
\beq
(-1)^{d_0+1 \o 2}   \frac{d F_{d_0 = {\rm  odd}}}{d \log r} &=& - \frac{  2 \pi^{d_0+1}  }{ d_0 !  } \sum_{i} c_i \B_i^2 +\cdots   \label{eq:oddflow} \\
(-1)^{\frac{d_0}{2}} \frac{d^2 F_{d_0 = {\rm  even}}}{d (\log r)^2} &=& - \frac{4 \pi^{d_0}}{d_0 !} \sum_{i} c_i \B_i^2 +\cdots .  \label{eq:evenflow}
\eeq
As is usual in perturbation theory, the higher order terms contain powers of the logarithm $\log (\m r)$ and we may set 
the renormalization scale as $\m \to r^{-1}$ to avoid large logarithmic corrections in perturbation theory. Then the coupling constants $\l (\mu)$
become functions of $r$ as $\l(\m) \to \l(r^{-1})$. 

Eqs.~(\ref{eq:oddflow}) and (\ref{eq:evenflow}) are our main result.
Similar results were obtained in Refs.~\cite{Cardy:1988cwa,Klebanov:2011gs} when a CFT is deformed by slightly marginal operators.
Eqs.~(\ref{eq:oddflow}) and (\ref{eq:evenflow}) extend those results to theories with marginal interactions.
In particular, these equations show that $(-1)^{d_0+1 \o 2}F_{d_0 = {\rm  odd}}$ and $(-1)^{\frac{d_0}{2}} d F_{d_0 = {\rm  even}}/d \log r $
decreases monotonically as we increase the radius of the sphere $r$ because the coefficients $c_i$ are positive.

\subsection{c-theorems in various dimensions}
\paragraph{Two dimensions}  
In two dimensions, the trace anomaly in CFT is given as~\footnote{
We do not introduce an additional factor of $2\pi$ in the definition of the energy-momentum tensor which often appears in the literature of two dimensional CFT.
}
\beq
T^\m_\m=\frac{c}{24 \pi} R, \label{eq:twodimcentral}
\eeq
where $c$ is the central charge.
The derivative of the free energy with respect to the radius of the sphere, $r$, is given by the one-point function of $T^\m_\m$ on the sphere $S^2$, and hence
\beq
\frac{d F_{d_0 = 2}}{d \log r } =-\vev{\int d^2 x \sqrt{\hat{g}} T^\m_\m }_{S^2}=-\frac{c}{3}. \label{eq:relationfandc}
\eeq
This relation is valid for CFT.

The Zamolodchikov's $c$-function~\cite{Zamolodchikov:1986gt}  is defined as
\beq
c(|x|)=(2\pi)^2 \left[ 2 z^4 \vev{T_{zz}(x)T_{zz}(0) }-4z^2 x^2 \vev{T_{z\bar{z}}(x)T_{zz}(0) }-6 x^4 \vev{T_{z\bar{z}}(x)T_{z\bar{z}}(0) }  \right],
\eeq
where $z=x^1+ix^2$. The function $c(|x|)$ coincides with the central charge $c$ in Eq.~(\ref{eq:twodimcentral}) at conformal fixed points.
The flow of this function is given by 
\beq
\frac{\d c(|x|)}{\d \log |x|} &=&-6\pi^2 x^4\vev{T^\m_\m(x) T^\n_\n(0)} \nonumber \\
&=&-6\pi^2\sum_{i,j}\B_i \B_j  x^4\vev{{\cal O}_i(x) {\cal O}_j(0)}. \label{eq:cfunctionflow}
\eeq
At the leading order, the operator correlation functions are given in Eq.~(\ref{eq:operatorcorrelation}).
Therefore, by comparing Eqs.~(\ref{eq:evenflow}) and (\ref{eq:cfunctionflow}), we find
\beq
\frac{d^2 F_{d_0 = 2}(r) }{d (\log r)^2 } =-\frac{1}{3} \left. \frac{d c(|x|)}{d \log |x|} \right|_{|x|=r}.
\eeq
This is consistent with Eq.~(\ref{eq:relationfandc}). Although the two functions $d F_{d_0 = 2} / d \log r $ and -$c(r)/3$ need not precisely be the same
in non-CFT, the above result shows that they indeed coincide at the order of perturbation theory we are considering. In particular,
our formula (\ref{eq:evenflow}) correctly reproduces the difference of UV and IR central charges 
$c_{\rm UV}-c_{\rm IR}$ if the UV and IR theories are conformal.

\paragraph{Four dimensions} 
The case of four dimensions is similar to that of two dimensions. The trace anomaly in CFT is given as
\beq
T^\m_\m=\frac{1}{16\pi^2} \left( -aE_4+cW_{\m\n\r\s}W^{\m\n\r\s} \right),
\eeq
where $W_{\m\n\r\s}$ is the Weyl tensor and $E_4=R^{\m\n\r\s}R_{\m\n\r\s}-4R_{\m\n}R^{\m\n}+R^2$ is the Euler density.
Putting the theory on the sphere $S^4$, we obtain
\beq
\frac{d F_{d_0 = 4}}{d \log r } =-\vev{\int d^2 x \sqrt{\hat{g}} T^\m_\m }_{S^2}=4a. \label{eq:relationfanda}
\eeq

The change of $a$ as we vary some length scale $r$ is given by~\cite{Komargodski:2011xv,Luty:2012ww}\footnote{
More precisely, $a$ is defined by the dilaton forward scattering amplitude, and $r$ should be the inverse of the center-of-mass energy
in that scattering process. See Ref.~\cite{Luty:2012ww} for details. }
\beq
\frac{d a(r)}{d \log r}=-\frac{\pi^4}{24}\sum_i c_i \B_i^2. \label{eq:afunctionflow}
\eeq
Therefore, by comparing Eqs.~(\ref{eq:evenflow}) and (\ref{eq:afunctionflow}) we get
\beq
\frac{d^2 F_{d_0 = 4}(r) }{d (\log r)^2 } =4  \frac{d a(r)}{d \log r} .
\eeq
This relation is consistent with Eq.~(\ref{eq:relationfanda}).

\paragraph{Three dimensions}
In three dimensional ${\cal N}=2$ supersymmetric theories there are exact results for the partition 
functions~\cite{Kapustin:2009kz,Jafferis:2010un,Hama:2010av}.\footnote{See also Refs.~\cite{Klebanov:2011gs,Klebanov:2011td} for calculations of $F$
in non-supersymmetric theories.}
We check our result for a simple case by comparing it to the exact result.

Let us consider an ${\cal N}=2$ supersymmetric Chern-Simons matter theory with gauge group $\U(N_c)$. 
We introduce chiral superfields $(Q,\tilde{Q})$ which are in a representation of the gauge group, $R \oplus \bar{R}$, where
we take $R$ as $N_f$ copies of some irreducible representation $r$, i.e., $R=N_f \times r$.
The Chern-Simons level is denoted as $k$, and we take $k \gg N_c,~N_f$ so that perturbation theory is applicable.
We take the superpotential as
\beq
W= \frac{\l}{2} (\tilde{Q} T_a Q)(\tilde{Q} T_a Q)
\eeq
where $T_a$ are generators of the gauge group $\U(N_c)$ normalized as $\tr_{\rm fund} (T_a T_b)=\d_{ab}$ for a fundamental representation.
Without loss of generality we take $\l$ to be real and positive. 
See Ref.~\cite{Gaiotto:2007qi} for details of this theory.

The theory has ${\cal N}=2$ supersymmetry for a general value of the yukawa coupling constant $\l$,
and the supersymmetry is enhanced to ${\cal N}=3$ when $\l=4\pi/k$. The RG equation for $\l$
is given by~\cite{Gaiotto:2007qi}~\footnote{Note that our normalization of $T_a$ is different from Ref.~\cite{Gaiotto:2007qi}, and 
we have also corrected some errors in the beta function of Ref.~\cite{Gaiotto:2007qi}.}
\beq
\B_\l=\frac{d \l}{d  \log \m}=\frac{b_0}{16\pi^2} \l \left( \l^2-\left(\frac{4\pi}{k} \right)^2 \right)
\eeq
where $b_0=\frac{2}{\dim R} (\tr_R (T_a T_b)\tr_R (T_a T_b)+\tr_R(T_a T_b T_a T_b))$ and $\dim R$ is the dimension of the representation $R$.
Therefore, this model connects two different superconformal fixed points; $\l=0$ in the UV and $\l=4\pi/k$ in the IR.

Let us apply our formula~(\ref{eq:oddflow}) to this model.
First, we define the operator ${\cal O}_\l$ as
\beq
{\cal O}_\l= \frac{1}{2}  \left. (\tilde{Q} T_a Q)(\tilde{Q} T_a Q) \right|_{\theta^2}+{\rm h.c.},
\eeq
where $|_{\theta^2}$ means that we take the $\theta^2$ component of a chiral field.
The Lagrangian of this theory contains the interaction term $\l {\cal O}_\l$.
By computing the correlation function $\vev{{\cal O}_\l(x){\cal O}_\l(0)}$ at the leading order, we find
that the constant $c_\l$ defined as $\vev{{\cal O}_\l(x){\cal O}_\l(0)}=c_\l /x^{6}$ is given by
\beq
c_\l&=&\frac{6b_0 \dim R}{(4\pi)^4},
\eeq
where we have neglected $O(\e)$ corrections.
Then, the difference of the UV and IR free energies, $F_{{\rm UV}}=F_{d_0=3}(r \to 0)$ and $F_{{\rm IR}}=F_{d_0=3}(r \to \infty)$, 
is given as
\beq
F_{{\rm IR}}- F_{{\rm UV}} &=&-\frac{\pi^4}{3} \int^\infty_0 \frac{d r}{r} c_\l \B_\l^2 \nonumber \\
&=&\frac{\pi^4}{3} \int^{4\pi \o k}_0 d \l c_\l \B_\l \nonumber \\
&=&-\frac{\pi^2 b_0^2 \dim R}{2^5 k^4}.
\eeq

Now let us restrict our attention to the case that the gauge group is $\U(1)$ (i.e., $N_c=1$) and there are $N_f$ pairs of chiral fields $(Q, \tilde{Q})$
with charge $\pm 1$. In this case, the gauge group generator is $T=1_{N_f \times N_f}$, and we have $b_0=2(N_f+1)$ and $\dim R=N_f$.
Therefore, we obtain
\beq
F_{{\rm IR}}- F_{{\rm UV}} =-\frac{\pi^2 N_f(N_f+1)^2 }{8 k^4}.\label{eq:perturbativechernsimons}
\eeq
On the other hand, the exact partition function for the model is explicitly obtained in Ref.~\cite{Jafferis:2010un}.
Denoting the superconformal $R$-charge of $(Q,\tilde{Q})$ as $\Delta=\frac{1}{2}-a$, the real part of the free energy~\footnote{
The imaginary part is just an artifact of imaginary supergravity 
background; see Refs.~\cite{Festuccia:2011ws,Closset:2012vg,Closset:2012vp} for details.} is given as
\beq
{\rm Re}F(a)=\log (2^{N_f} \sqrt{k}) -\frac{\pi^2 N_f }{2} a^2+\frac{\pi^2 N_f(N_f+1)}{16 k^2} (1+8a) + O(k^{-5}),
\eeq
where we have assumed $a =O(k^{-2})$, which will be justified below. In the UV CFT ($\l=0$), the value of $a$ is determined by the solution of
$d {\rm Re}F(a)/ d a=0$~\cite{Jafferis:2010un,Closset:2012vg} 
and is given by $a=\frac{N_f+1}{2k^2}$. In the IR CFT ($\l=4\pi/k$), we should have $a=0$ for the superpotential 
to be invariant under the $R$-symmetry. Then, we obtain
\beq
{\rm Re}F_{\rm IR}-{\rm Re}F_{\rm UV} &=& {\rm Re}F(a=0)-{\rm Re}F(a=\frac{N_f+1}{2k^2}) \nonumber \\
&=&-\frac{\pi^2 N_f(N_f+1)^2}{8k^4}.
\eeq
This result completely agrees with Eq.~(\ref{eq:perturbativechernsimons}).

\paragraph{Six dimensions}
In six dimensions, scalar field theories with $\phi^3$ interactions can be treated perturbatively. The Lagrangian is given by 
\beq
{\cal L}=\frac{1}{2} \sum_a \q_\m \phi_a \q^\m \phi_a+\frac{1}{6} \sum_{a,b,c} \l_{a,b,c}\phi_a \phi_b \phi_c.
\eeq
Our result shows that $-d F_{d_0=6} / d \log r$ decreases monotonically as we increase $r$.
Notice that the conformal coupling $\frac{d-2}{8(d-1)} R \phi^2$ makes the vacuum perturbatively stable when the theory is put on the sphere,
and hence we need not worry about infrared divergences.

This model is asymptotically free at the one-loop level~\cite{Macfarlane:1974vp},
but unfortunately, no Banks-Zaks type IR fixed point is known.
However, it is at least encouraging for the six dimensional $a$-theorem~\cite{Elvang:2012st} (see also Ref.\cite{Elvang:2012yc})
that $d F_{d_0=6} / d \log r$ decreases monotonically as a function of $r$.

\subsection{Scale versus conformal invariance}
Now we discuss the equivalence between scale and conformal invariance in the class of theories studied in this paper (see Table.~\ref{table:1}).
More generally, we study the possible IR (or UV) asymptotics of perturbative quantum field theories.
Our discussion follows the one in Ref.~\cite{Luty:2012ww} which studied the same problem in four dimensions.

First let us briefly review the mechanism by which a theory could have scale invariance 
without conformal invariance~\cite{Fortin:2011ks,Fortin:2011sz,Fortin:2012ic,Fortin:2012cq}.
If the trace of the energy-momentum tensor is given by a total derivative, $T^\m_\m=\q_\m j^\m_V$,
where $j_V^\m$ is some vector field called the virial current, the theory is scale invariant
because we can define a conserved current of scaling transformation as $T^{\m\n}x_\n-j_V^\m$.
In theories studied in this paper, this requirement is given as
\beq
T^\m_\m=-\sum_i \B^i [{\cal O}_i]=\q_\m j_V^\m. \label{eq:scaleinvtrace}
\eeq
If $\q_\m  j_V^\m \neq 0$, the beta functions $\B^i$ are nonzero and the theory is not conformally invariant.
If we see the coupling constants of the theory as spurions, 
there is a symmetry associated with the current $j_V$ under which the coupling constants transform nontrivially.
When Eq.~(\ref{eq:scaleinvtrace}) holds, the RG flow is generated by that symmetry acting on the coupling constants.
Our purpose is to show that  such RG flows are impossible and $T^\m_\m$ actually vanishes when the theory has scale invariance.

The free energy $F$ in general depends on the parameters in the finite part of the counterterms 
$S_{\rm c.t.}[\hat{g}_{\m\n}]$. These new parameters are absent in the original flat space theories.
However, as we discussed in section~\ref{sec:2}, 
there is a way to define the free energy in which $F$ in odd dimensions and $d F/ d\log r$ in even dimensions do not contain
any such new parameters.
In that definition, they only depend on the coupling constants $\l_i(\mu)$, the renormalization scale $\m$, and the radius of the sphere $r$ as
\beq
C &=& C \left(\l(\mu), \log(r/\mu) \right) =C(\l(r^{-1})) \label{eq:functionalform}
\eeq
where we define $C$ as
\beq
C= \left\{
\begin{array}{ll}
(-1)^{d_0+1 \o 2} \displaystyle{\frac{ d_0 !  }{  2 \pi^{d_0+1}  }} F_{d_0 = {\rm  odd}} & (d_0={\rm odd})  \\
\\
(-1)^{\frac{d_0}{2}} \displaystyle{ \frac{d_0 !}{4 \pi^{d_0}}} \displaystyle{ \frac{d F_{d_0 = {\rm  even}}}{d \log r}}  & ( d_0={\rm even}) 
\end{array}
\right. . \label{eq:defineCfunction}
\eeq
In the last equality in Eq.~(\ref{eq:functionalform}) we have used RG invariance and set $\mu=r^{-1}$.
Therefore, the function $C$ defined as in Eq.~(\ref{eq:defineCfunction}) is only a function of $\l_i(r^{-1})$.

From Eqs.~(\ref{eq:oddflow}) and (\ref{eq:evenflow}), we see that $C$ satisfies 
\beq
{d C \o d \log r}=-\sum_i c_i \B_i^2 \label{eq:Cfloweq}
\eeq 
with $c_i$ all positive. 
Suppose that the theory is weakly coupled in the IR limit $r \to \infty$. (The discussion is completely parallel in the UV.)
Then, we can trust perturbation theory in the IR, and $C(\l(r^{-1}))$ remains finite in the IR since all the couplings $\l_i(r^{-1})$ are small. 
Then, from Eq.~(\ref{eq:Cfloweq}), it is necessary that $\int ^\infty d\log r  \sum_i c_i \B_i^2$ is finite, and hence
$c_i \B_i^2$ should vanish faster than $1/\log r$ in the limit $r \to \infty$
for $C$ to be finite in the IR. Since $\vev{T^\m_\m(x) T^\n_\n(0)} =\sum_i c_i \B_i^2 /x^{2d} $,  the trace of the energy-momentum
tensor should vanish in the IR limit and hence the theory is conformal. We conclude that the IR limit of the class of theories studied in this paper
is either conformal or strongly coupled so that perturbation breaks down.

Although we have neglected higher order corrections, they do not change the conclusion.
As long as the energy-momentum tensor satisfies
Eq.~(\ref{eq:curvedtrace}), couplings between
the dilaton and dynamical fields in Eq.~(\ref{eq:mattertau}) are always proportional to the beta functions $\B_i$.
Then, higher order corrections to Eq.~(\ref{eq:Cfloweq}) are of the form $ \sum_{i,j} \d c_{ij} \B_i \B_j$, where $\d c_{ij}$
are suppressed by loop factors compared with $c_i$. These corrections are always smaller than the leading contribution and
the above discussion is valid even if we include them. See Ref.~\cite{Luty:2012ww} for detailed discussions.
Furthermore, the ambiguity of the energy-momentum tensor related to improvement discussed in section~\ref{sec:2}
does not invalidate the above argument if there exists a solution to Eq.~(\ref{eq:additionalparameter})
in which $\xi^a_i$ remains small in the IR. In particular, in a scale invariant theory in which the energy-momentum tensor
and the virial current $j^\m_V$ in Eq.~(\ref{eq:scaleinvtrace}) are eigenstates of dilatation with eigenvalues $d$ and $d-1$ 
respectively,
there is no operator mixing and the above argument of the equivalence between scale and conformal invariance is valid.

\section{Conclusions}\label{sec:4}
In this paper we have studied the free energy $F=-\log Z$ on a $d$-dimensional sphere with radius $r$ for theories which have marginal
interactions. Such theories are listed in Table.~\ref{table:1}.  If we couple the metric of the sphere to the energy-momentum tensor $T_{\m\n}$
satisfying $T^\m_\m=-\sum_i \B^i [{\cal O}_i]$,
the free energy $F$ satisfies 
\beq
(-1)^{d_0+1 \o 2}   \frac{d F_{d_0 = {\rm  odd}}}{d \log r} &=& - \frac{  2 \pi^{d_0+1}  }{ d_0 !  } \sum_{i} c_i \B_i^2 +\cdots   \label{eq:concloddflow} \\
(-1)^{\frac{d_0}{2}} \frac{d^2 F_{d_0 = {\rm  even}}}{d (\log r)^2} &=& - \frac{4 \pi^{d_0}}{d_0 !} \sum_{i} c_i \B_i^2 +\cdots ,  \label{eq:conclevenflow}
\eeq
where $d_0$ is the space-time dimension, $\B_i$ are beta functions of coupling constants, 
$c_i$ are positive constants defined as $\vev{[{\cal O}_i(x)][{\cal O}_j(0)]}=c_i \d_{ij} / x^{2d}+\cdots$, 
and dots indicate sub-leading terms which are always smaller than the leading term.
In particular, $(-1)^{\frac{d_0+1}{2}} F$ (in odd dimensions) or $(-1)^{\frac{d_0}{2}} d F / d \log r$ (in even dimensions) decreases monotonically
in perturbation theory.
This result extends the perturbative $c$ ($a$) theorem in two (four) dimensions to other dimensions. 
Using this result, we have extended the work of Ref.~\cite{Luty:2012ww,Fortin:2012hn} to other dimensions
and argued that scale invariance is equivalent to conformal invariance in
perturbation theory.

\section*{Acknowledgements}
The author would like to thank N.~Kim, I.~Klebanov, J.~Maldacena, B.~Safdi, E.~Witten and especially T.~Nishioka and M.~Yamazaki 
for useful discussions. The work of KY is supported by NSF grant PHY-0969448.


\begin{thebibliography}{99}

\bibitem{Zamolodchikov:1986gt} 
  A.~B.~Zamolodchikov,
  ``Irreversibility of the Flux of the Renormalization Group in a 2D Field Theory,''
  JETP Lett.\  {\bf 43}, 730 (1986)
  [Pisma Zh.\ Eksp.\ Teor.\ Fiz.\  {\bf 43}, 565 (1986)].
  
\bibitem{Polchinski:1987dy} 
  J.~Polchinski,
  ``Scale And Conformal Invariance In Quantum Field Theory,''
  Nucl.\ Phys.\ B {\bf 303}, 226 (1988).
  


  
  
\bibitem{Deser:1993yx} 
  S.~Deser and A.~Schwimmer,
  ``Geometric classification of conformal anomalies in arbitrary dimensions,''
  Phys.\ Lett.\ B {\bf 309}, 279 (1993)
  [hep-th/9302047].
  
\bibitem{Cardy:1988cwa} 
  J.~L.~Cardy,
  ``Is There a c Theorem in Four-Dimensions?,''
  Phys.\ Lett.\ B {\bf 215}, 749 (1988).

  
\bibitem{Jafferis:2011zi} 
  D.~L.~Jafferis, I.~R.~Klebanov, S.~S.~Pufu and B.~R.~Safdi,
  ``Towards the F-Theorem: N=2 Field Theories on the Three-Sphere,''
  JHEP {\bf 1106}, 102 (2011)
  [arXiv:1103.1181 [hep-th]].
  
\bibitem{Klebanov:2011gs} 
  I.~R.~Klebanov, S.~S.~Pufu and B.~R.~Safdi,
  ``F-Theorem without Supersymmetry,''
  JHEP {\bf 1110}, 038 (2011)
  [arXiv:1105.4598 [hep-th]].

\bibitem{Komargodski:2011vj} 
  Z.~Komargodski and A.~Schwimmer,
  ``On Renormalization Group Flows in Four Dimensions,''
  JHEP {\bf 1112}, 099 (2011)
  [arXiv:1107.3987 [hep-th]].
  
  
  
\bibitem{Komargodski:2011xv} 
  Z.~Komargodski,
  ``The Constraints of Conformal Symmetry on RG Flows,''
  JHEP {\bf 1207}, 069 (2012)
  [arXiv:1112.4538 [hep-th]].

\bibitem{Luty:2012ww} 
  M.~A.~Luty, J.~Polchinski and R.~Rattazzi,
 ``The $a$-theorem and the Asymptotics of 4D Quantum Field Theory,''
  arXiv:1204.5221 [hep-th].

  
  
  
\bibitem{Casini:2012ei} 
  H.~Casini and M.~Huerta,
  ``On the RG running of the entanglement entropy of a circle,''
  Phys.\ Rev.\ D {\bf 85}, 125016 (2012)
  [arXiv:1202.5650 [hep-th]].
  
\bibitem{Casini:2011kv} 
  H.~Casini, M.~Huerta and R.~C.~Myers,
  ``Towards a derivation of holographic entanglement entropy,''
  JHEP {\bf 1105}, 036 (2011)
  [arXiv:1102.0440 [hep-th]].
  
  
  
\bibitem{Freedman:1999gp} 
  D.~Z.~Freedman, S.~S.~Gubser, K.~Pilch and N.~P.~Warner,
  ``Renormalization group flows from holography supersymmetry and a c theorem,''
  Adv.\ Theor.\ Math.\ Phys.\  {\bf 3}, 363 (1999)
  [hep-th/9904017].
  
\bibitem{Myers:2010xs} 
  R.~C.~Myers and A.~Sinha,
  ``Seeing a c-theorem with holography,''
  Phys.\ Rev.\ D {\bf 82}, 046006 (2010)
  [arXiv:1006.1263 [hep-th]].
  
\bibitem{Myers:2010tj} 
  R.~C.~Myers and A.~Sinha,
  ``Holographic c-theorems in arbitrary dimensions,''
  JHEP {\bf 1101}, 125 (2011)
  [arXiv:1011.5819 [hep-th]].
  
  
\bibitem{Elvang:2012st} 
  H.~Elvang, D.~Z.~Freedman, L.~-Y.~Hung, M.~Kiermaier, R.~C.~Myers and S.~Theisen,
  ``On renormalization group flows and the a-theorem in 6d,''
  JHEP {\bf 1210}, 011 (2012)
  [arXiv:1205.3994 [hep-th]].
  
\bibitem{Elvang:2012yc} 
  H.~Elvang and T.~M.~Olson,
  ``RG flows in d dimensions, the dilaton effective action, and the a-theorem,''
  arXiv:1209.3424 [hep-th].
  
\bibitem{Bhattacharyya:2012tc} 
  A.~Bhattacharyya, L.~-Y.~Hung, K.~Sen and A.~Sinha,
  ``On c-theorems in arbitrary dimensions,''
  arXiv:1207.2333 [hep-th].
  
\bibitem{Fortin:2012hn} 
  J.~-F.~Fortin, B.~Grinstein and A.~Stergiou,
  ``Limit Cycles and Conformal Invariance,''
  arXiv:1208.3674 [hep-th].

  
   
\bibitem{Callan:1970ze} 
  C.~G.~Callan, Jr., S.~R.~Coleman and R.~Jackiw,
  ``A New improved energy - momentum tensor,''
  Annals Phys.\  {\bf 59}, 42 (1970).
  
\bibitem{Dorigoni:2009ra} 
  D.~Dorigoni and V.~S.~Rychkov,
  ``Scale Invariance + Unitarity => Conformal Invariance?,''
  arXiv:0910.1087 [hep-th].
  
\bibitem{Antoniadis:2011gn} 
  I.~Antoniadis and M.~Buican,
  ``On R-symmetric Fixed Points and Superconformality,''
  Phys.\ Rev.\ D {\bf 83}, 105011 (2011)
  [arXiv:1102.2294 [hep-th]].
  
\bibitem{Zheng:2011bp} 
  S.~Zheng and Y.~Yu,
  ``Is There Scale Invariance in N=1 Supersymmetric Field Theories ?,''
  arXiv:1103.3948 [hep-th].
  
  
  
\bibitem{Fortin:2011ks} 
  J.~-F.~Fortin, B.~Grinstein and A.~Stergiou,
  ``Scale without Conformal Invariance: An Example,''
  Phys.\ Lett.\ B {\bf 704}, 74 (2011)
  [arXiv:1106.2540 [hep-th]].
  
\bibitem{Fortin:2011sz} 
  J.~-F.~Fortin, B.~Grinstein and A.~Stergiou,
  ``Scale without Conformal Invariance: Theoretical Foundations,''
  JHEP {\bf 1207}, 025 (2012)
  [arXiv:1107.3840 [hep-th]].
  
\bibitem{Fortin:2012ic} 
  J.~-F.~Fortin, B.~Grinstein and A.~Stergiou,
  ``Scale without Conformal Invariance at Three Loops,''
  JHEP {\bf 1208}, 085 (2012)
  [arXiv:1202.4757 [hep-th]].
  
\bibitem{Fortin:2012cq} 
  J.~-F.~Fortin, B.~Grinstein and A.~Stergiou,
  ``Scale without Conformal Invariance in Four Dimensions,''
  arXiv:1206.2921 [hep-th].
  
  
    
%
  
  
\bibitem{Nakayama:2011tk} 
  Y.~Nakayama,
  ``Comments on scale invariant but non-conformal supersymmetric field theories,''
  Int.\ J.\ Mod.\ Phys.\ A {\bf 27}, 1250122 (2012)
  [arXiv:1109.5883 [hep-th]].
  
  
\bibitem{Nakayama:2012nd} 
  Y.~Nakayama,
  ``Supercurrent, Supervirial and Superimprovement,''
  arXiv:1208.4726 [hep-th].
  
\bibitem{Jack:1990eb} 
  I.~Jack and H.~Osborn,
  ``Analogs For The C Theorem For Four-dimensional Renormalizable Field Theories,''
  Nucl.\ Phys.\ B {\bf 343}, 647 (1990).
  
\bibitem{Osborn:1991gm} 
  H.~Osborn,
  ``Weyl consistency conditions and a local renormalization group equation for general renormalizable field theories,''
  Nucl.\ Phys.\ B {\bf 363}, 486 (1991).
  

  
\bibitem{ElShowk:2011gz} 
  S.~El-Showk, Y.~Nakayama and S.~Rychkov,
  ``What Maxwell Theory in D<>4 teaches us about scale and conformal invariance,''
  Nucl.\ Phys.\ B {\bf 848}, 578 (2011)
  [arXiv:1101.5385 [hep-th]].
  
\bibitem{Jackiw:2011vz} 
  R.~Jackiw and S.~-Y.~Pi,
  ``Tutorial on Scale and Conformal Symmetries in Diverse Dimensions,''
  J.\ Phys.\ A {\bf 44}, 223001 (2011)
  [arXiv:1101.4886 [math-ph]].
  


\bibitem{Weinberg:1996kr} 
  S.~Weinberg,
  ``The quantum theory of fields. Vol. 2: Modern applications,''
  Cambridge, UK: Univ. Pr. (1996) 489 p
  
\bibitem{Collins:1976vm} 
  J.~C.~Collins,
  ``The Energy-Momentum Tensor Revisited,''
  Phys.\ Rev.\ D {\bf 14}, 1965 (1976).
  

  


\bibitem{Schwimmer:2010za} 
  A.~Schwimmer and S.~Theisen,
  ``Spontaneous Breaking of Conformal Invariance and Trace Anomaly Matching,''
  Nucl.\ Phys.\ B {\bf 847}, 590 (2011)
  [arXiv:1011.0696 [hep-th]].

  
\bibitem{Duff:1993wm} 
  M.~J.~Duff,
  ``Twenty years of the Weyl anomaly,''
  Class.\ Quant.\ Grav.\  {\bf 11}, 1387 (1994)
  [hep-th/9308075].
  
  
    
\bibitem{Ferrara:1974pz} 
  S.~Ferrara and B.~Zumino,
  ``Transformation Properties of the Supercurrent,''
  Nucl.\ Phys.\ B {\bf 87}, 207 (1975).
  
\bibitem{Komargodski:2009pc} 
  Z.~Komargodski and N.~Seiberg,
  ``Comments on the Fayet-Iliopoulos Term in Field Theory and Supergravity,''
  JHEP {\bf 0906}, 007 (2009)
  [arXiv:0904.1159 [hep-th]].
  
\bibitem{Komargodski:2010rb} 
  Z.~Komargodski and N.~Seiberg,
  ``Comments on Supercurrent Multiplets, Supersymmetric Field Theories and Supergravity,''
  JHEP {\bf 1007}, 017 (2010)
  [arXiv:1002.2228 [hep-th]].
  
\bibitem{Dumitrescu:2011iu} 
  T.~T.~Dumitrescu and N.~Seiberg,
  ``Supercurrents and Brane Currents in Diverse Dimensions,''
  JHEP {\bf 1107}, 095 (2011)
  [arXiv:1106.0031 [hep-th]].
  
  
  
  
  
\bibitem{Clark:1978jx} 
  T.~E.~Clark, O.~Piguet and K.~Sibold,
  ``Supercurrents, Renormalization and Anomalies,''
  Nucl.\ Phys.\ B {\bf 143}, 445 (1978).
  
\bibitem{Piguet:1981mu} 
  O.~Piguet and K.~Sibold,
  ``The Supercurrent In N=1 Supersymmetrical Yang-mills Theories. 1. The Classical Case,''
  Nucl.\ Phys.\ B {\bf 196}, 428 (1982).
  
\bibitem{Piguet:1981mw} 
  O.~Piguet and K.~Sibold,
  ``The Supercurrent In N=1 Supersymmetrical Yang-mills Theories. 2. Renormalization,''
  Nucl.\ Phys.\ B {\bf 196}, 447 (1982).
  
  
  
\bibitem{Grisaru:1978vx} 
  M.~T.~Grisaru,
  ``Anomalies In Supersymmetric Theories,''
  In *Shifman, M.A. (ed.): The many faces of the superworld* 370-387
  
  
  
\bibitem{Grisaru:1985ik} 
  M.~T.~Grisaru, B.~Milewski and D.~Zanon,
  ``The Supercurrent And The Adler-bardeen Theorem,''
  Nucl.\ Phys.\ B {\bf 266}, 589 (1986).
  
\bibitem{Shifman:1986zi} 
  M.~A.~Shifman and A.~I.~Vainshtein,
  ``Solution of the Anomaly Puzzle in SUSY Gauge Theories and the Wilson Operator Expansion,''
  Nucl.\ Phys.\ B {\bf 277}, 456 (1986)
  [Sov.\ Phys.\ JETP {\bf 64}, 428 (1986)]
  [Zh.\ Eksp.\ Teor.\ Fiz.\  {\bf 91}, 723 (1986)].
  
  
  
  
\bibitem{ArkaniHamed:1997mj} 
  N.~Arkani-Hamed and H.~Murayama,
 ``Holomorphy, rescaling anomalies and exact beta functions in supersymmetric gauge theories,''
  JHEP {\bf 0006}, 030 (2000)
  [hep-th/9707133].
  
  
\bibitem{Huang:2010tn} 
  X.~Huang and L.~Parker,
  ``Clarifying Some Remaining Questions in the Anomaly Puzzle,''
  Eur.\ Phys.\ J.\ C {\bf 71}, 1570 (2011)
  [arXiv:1001.2364 [hep-th]].
  
\bibitem{Yonekura:2010mc} 
  K.~Yonekura,
  ``Notes on Operator Equations of Supercurrent Multiplets and Anomaly Puzzle in Supersymmetric Field Theories,''
  JHEP {\bf 1009}, 049 (2010)
  [arXiv:1004.1296 [hep-th]].
  
\bibitem{Yonekura:2012uk} 
  K.~Yonekura,
  ``On the Trace Anomaly and the Anomaly Puzzle in N=1 Pure Yang-Mills,''
  JHEP {\bf 1203}, 029 (2012)
  [arXiv:1202.1514 [hep-th]].
  
  
  
  
\bibitem{Kapustin:2009kz} 
  A.~Kapustin, B.~Willett and I.~Yaakov,
  ``Exact Results for Wilson Loops in Superconformal Chern-Simons Theories with Matter,''
  JHEP {\bf 1003}, 089 (2010)
  [arXiv:0909.4559 [hep-th]].
  
\bibitem{Jafferis:2010un} 
  D.~L.~Jafferis,
  ``The Exact Superconformal R-Symmetry Extremizes Z,''
  JHEP {\bf 1205}, 159 (2012)
  [arXiv:1012.3210 [hep-th]].
  
\bibitem{Hama:2010av} 
  N.~Hama, K.~Hosomichi and S.~Lee,
  ``Notes on SUSY Gauge Theories on Three-Sphere,''
  JHEP {\bf 1103}, 127 (2011)
  [arXiv:1012.3512 [hep-th]].
  
\bibitem{Klebanov:2011td} 
  I.~R.~Klebanov, S.~S.~Pufu, S.~Sachdev and B.~R.~Safdi,
  ``Entanglement Entropy of 3-d Conformal Gauge Theories with Many Flavors,''
  JHEP {\bf 1205}, 036 (2012)
  [arXiv:1112.5342 [hep-th]].


\bibitem{Gaiotto:2007qi} 
  D.~Gaiotto and X.~Yin,
  ``Notes on superconformal Chern-Simons-Matter theories,''
  JHEP {\bf 0708}, 056 (2007)
  [arXiv:0704.3740 [hep-th]].
  
\bibitem{Closset:2012vg} 
  C.~Closset, T.~T.~Dumitrescu, G.~Festuccia, Z.~Komargodski and N.~Seiberg,
  ``Contact Terms, Unitarity, and F-Maximization in Three-Dimensional Superconformal Theories,''
  JHEP {\bf 1210}, 053 (2012)
  [arXiv:1205.4142 [hep-th]].
  
  
\bibitem{Festuccia:2011ws} 
  G.~Festuccia and N.~Seiberg,
  ``Rigid Supersymmetric Theories in Curved Superspace,''
  JHEP {\bf 1106}, 114 (2011)
  [arXiv:1105.0689 [hep-th]].
  
  
\bibitem{Closset:2012vp} 
  C.~Closset, T.~T.~Dumitrescu, G.~Festuccia, Z.~Komargodski and N.~Seiberg,
  ``Comments on Chern-Simons Contact Terms in Three Dimensions,''
  JHEP {\bf 1209}, 091 (2012)
  [arXiv:1206.5218 [hep-th]].
  
\bibitem{Macfarlane:1974vp} 
  A.~J.~Macfarlane and G.~Woo,
  ``Phi**3 Theory in Six-Dimensions and the Renormalization Group,''
  Nucl.\ Phys.\ B {\bf 77}, 91 (1974).
  
  
    


\end{thebibliography}
\end{document}